\documentclass[12pt]{article}
\usepackage{graphicx}
\def\beq{\begin{equation}}
\def\eeq{\end{equation}}
\def\bea{\begin{eqnarray}}
\def\eea{\end{eqnarray}}

\def\roughly#1{\mathrel{\raise.3ex\hbox
{$#1$\kern-.75em\lower1ex\hbox{$\sim$}}}}

\def\sss{\scriptscriptstyle}

\def\CA{{\cal A}}

\def\barpk{{\raise.35ex\hbox  {${\sss  (}$}}--{\raise.35ex\hbox{${\sss
)}$}}}        \def\bbarp{\hbox{$B$\kern-0.9em\raise1.4ex\hbox{\barpk}}}

  \def\rr2{{1\over\sqrt{2}}}
\def\wt{\widetilde}
\def\.{\!\cdot\!}    \def\:{\cdots}   \def\[{\left[}   \def\]{\right]}
\def\({\left(} \def\){\right)} 
\def\rr2{{1\over\sqrt{2}}}
\def\tcV{ t \to c V, V=\gamma, Z}
\def\bsV{ b \to s V, V=\gamma, Z}
\def\tcZ{ t \to c Z}
\def\tcgam{ t \to c \gamma}
\def\bsZ{ b \to s Z}

\begin{document}
\begin{flushright}
UMISS-HEP-2009-02 \\
[10mm]
\end{flushright}

\begin{center}
\bigskip {\Large  \bf Model Independent Predictions for Rare Top Decays 
with Weak Coupling}
\\[8mm]
Alakabha Datta and Murugeswaran Duraisamy
\footnote{E-mail:
\texttt{datta@phy.olemiss.edu}} 
\footnote{E-mail:
\texttt{duraism@phy.olemiss.edu}}  
\\[3mm]
\end{center}

\begin{center}
~~~~~~~~~~~ {\it Department  of Physics and Astronomy,}\\ 
~~~~~~~~~~~~{ \it University of Mississippi,}\\
~~~~~~~~~~~~{\it  Lewis Hall, University, MS, 38677.}\\
\end{center}


\begin{center} 
\bigskip (\today) \vskip0.5cm {\Large Abstract\\} \vskip3truemm
\parbox[t]{\textwidth}  {Measurements at  $B$ factories  have provided
important  constraints  on  new  physics  in  several  rare  processes
involving the  $B$ meson.   New Physics, if  present in the  $b$ quark
sector may  also affect  the top sector.   In an  effective Lagrangian
approach, we write down operators  where effects in the bottom and the
top sector are related. Assuming  the couplings of the operators to be
of the  same size as the weak  coupling $g$ of the  Standard Model and
taking into account constraints on  new physics from the bottom sector as 
well as top branching ratios,
we  make  predictions for  the  rare  top decays  $  t  \to cV$  where
$V=\gamma,  Z$. We find  branching fractions  for these  decays within
possible reach of the LHC. Predictions  are also made for $t \to s W$.
}
\end{center}

\thispagestyle{empty} \newpage \setcounter{page}{1}
\baselineskip=14pt

\section{Introduction}
The flavor sector of the  Standard Model(SM) is poorly understood. The
origin of masses  and mixing and CP violation in  the quark and lepton
sector  is  unknown. Another  mystery is  the  rare Flavor  Changing
Neutral  Current processes.   Flavor Changing  Neutral  Current (FCNC)
processes in the Standard Model(SM) do not arise at tree level, and are
highly  suppressed. Many  extensions  of the  SM  naturally have  FCNC
processes that  occur at tree  or loop level.  Hence,  measurements of
FCNC processes can  put strong constraints on new  physics (NP) that may be
discovered at present colliders like the Tevatron or the LHC.  In that
sense, flavor data can complement the new physics search at colliders.
 
Effects  from  new  physics  can  cause  deviations  from  the  SM
predictions.  These  deviations are expected to be  more pronounced in
rare FCNC processes as they are suppressed in the SM.  The B factories
have made several measurements of  FCNC processes in the bottom sector
and  have  put strong constraints on new physics.  Here we will be
concerned with constraints  on the $b \to  s \gamma$ and $ b  \to s Z$
transitions.  New physics  in  the former  are  constrained by  better
measurements of  the $b  \to s \gamma$  rate \cite{hfag} and  a better
understanding of the SM \cite{misiak} contribution to
the process. The later transition  is constrained by $B_s$ mixing, $ b
\to s ll$ and also possible hints of new physics in decays like $B \to
K \pi, \phi K_s$ etc \cite{ZFCNC,ZFCNC2}.

There are no  measurements of FCNC in the top  sector.  There are 95\%
C.L bounds,  $ t \to q(=u,c) \gamma  <5.9 \times 10^{-3}$ and  $ t \to
q(=u,c) Z <0.037$ \cite{pdg}.  In  the SM the branching ratios for the
rare  FCNC  decays  $  t  \to  cV$  where  $V=g,\gamma,  Z$  are  tiny
\cite{soni,mele}. The small mass of  the internal quarks in the SM loop
diagram makes  FCNC effects in the  top sector much  smaller than FCNC
effects in the bottom sector.  Hence, FCNC processes in the top sector
are excellent probes of new physics.

The LHC will be a top  factory allowing the possible detection of FCNC
effects in the top sector \cite{lhcfcnc}.  One can hope to measure $ t
\to q (=u,c)Z$ with branching  ratios in the range $6.1 \times 10^{-5}
- 3.1 \times  10^{-4}$ while $ t  \to q (=u,c)\gamma$  can be measured
with branching  ratios in the range  $1.2 \times 10^{-5}  - 4.1 \times
10^{-5}$.  New physics  searches via  the  top quark  decays have  been
extensively   analyzed   in   the   literature  in   specific   models
\cite{topfcnc}.  In this  paper we focus on a  model independent study
of the non-SM FCNC effects in the top sector. In this framework, imposing
the constraints  on $\bsV$  transitions as well as constraints from top branching ratios measurements, we predict  the size  of rare
FCNC $\tcV$ decays.

In our  approach, we write  down higher dimension operators  which are
invariant under the SM gauge group that generate the anomalous $\tcV$
couplings. As  the left-handed top  and the left-handed bottom  are in
the same  $SU(2)_{L}$ doublet  the $tcV$ and  the $bsV$  couplings are
related.  We  consider  two operators that  can generate the
$tcV$ and $bsV$ couplings. One involves the $SU(2)_L$ gauge fields and
the  other  the  $U(1)_Y$ gauge  field.  We  choose  the size  of  the
couplings to  be the same size  as the $SU(2)_L$  gauge coupling, $g$,
and the $U(1)_Y$ gauge coupling $g'$.  This choice for the size of the
anomalous  coupling is motivated  by the  assumption that  the physics
that   generates   the  anomalous   couplings   are  weakly   coupled.
Constraints from $b \to s  \gamma$ force the couplings between the two
operators  to follow  the same  relation as  the   one  between the
$U(1)_Y$  and $SU(2)_L$  gauge  couplings in  the  SM to  a very  good
approximation.  Assuming  such a  relation between the  two couplings,
the $ b \to s \gamma$ constraint is eliminated and all predictions are
found to  depend on  a single coupling  associated with  the $SU(2)_L$
gauge field. With  the size of this coupling of the  same order as $g$,
all low  energy constraints are  found to be satisfied.   The operators
also generate a  $ t \to s  W$ vertex and for the  anomalous coupling $
\sim g$, the corrections to the branching  ratio for $ t \to s W$ from
new physics is found to be consistent with 
the top branching fraction measurements. Finally, predictions are made for $
t \to c \gamma, t \to c Z$ and $ t \to s W$ transitions.

There  have  been  previous  attempts  \cite{han1,han2,han3}  to  make
predictions for rare top  processes using constraints from $B$ decays,
specifically   $  b  \to   s  \gamma$,   in  an   effective  Lagrangian
approach.  There are  several differences  between this  work  and the
previous work.  First,  in the previous work the  $tc\gamma$ and $tcZ$
couplings are  independent while in our  work they are  related as our
anomalous couplings  are generated by operators invariant  under the SM
gauge group. Second, in the previous work constraints on the anomalous
$tcZ$ and $tc \gamma$ couplings  are obtained from FCNC effects in the
down  sector generated  though loop  effects.   In our  work, for  the
considered  size of  the  couplings,  we find  that  loop effects  are
sufficiently small to be  consistent with experiments and therefore do
not introduce  any additional constraints.  The size of  the anomalous
couplings are fixed from tree  processes and hence these couplings are
quite strongly constrained. As indicated above, we also  take into 
account experimental 
constraints on top branching fractions.

  Finally, a unique feature of the operators in
the effective  Lagrangian in  our approach is  that they  are momentum
dependent  and therefore  contributions  to FCNC  effects  in the  top
sector are enhanced typically by  a factor $ \sim {m_t^2 \over m_b^2}$
relative to  the ones  in the  bottom sector. Note  that, it  has been
speculated in  the past that  FCNC effects in the  top sector may  be enhanced
because of its heavy mass.  This has motivated specific ansatz for the
FCNC vertices with enhanced effect in the top sector\cite{sher}.

In our approach, the anomalous  couplings in the bottom and top sector
are  related.   This is  true  only  for  certain classes  of  models.
However,  the connection  between the  top and  bottom sectors  is not
generic as far as FCNC effects are concerned. In the two higgs doublet
model, for  instance, FCNC arise in  the bottom and the  top sector at
the tree or loop level.  However, any connection between the effects in
the two sectors are strongly  dependent on the structure of the Yukawa
couplings  in the  up and  the  down quark  sectors.  Within  specific
models of the Yukawa structures one can relate FCNC effects in the top
and the bottom sector in new physics models \cite{model23, sfs, valencia}.

The paper is  organized in the following manner.  In  sec.~1 we write
down the effective Hamiltonian that  generates the $t \to c V(=\gamma,
Z)$ transitions. The vertices for $b \to  s V, t \to cV$ as well as $b
\to  c W$  and $  t \to  s W$  are written  down. Constraints  on these
couplings  are  obtained.  In  the   next  section,  sec.~2,  we  make
predictions for the processes $  t \to c V$, $  V=\gamma, Z$, and 
$ t \to  s W$.  In
the  final   section, we present  our
conclusions.

\section{ Effective Lagrangian}
In this section we write  the effective Lagrangian that generates to $
\tcV$ transitions.  We write  the effective Hamiltonian as, \bea {\cal
L}& = & {\cal L}_{SM} + \sum_i\frac{c_i {\cal O}_i}{\Lambda^2}, \
\label{eff_lag}
\eea where ${\cal O}_i$ are  dimension 6 operators.

We will  concentrate on the following two  operators \cite{buch}, \bea
{\cal O}_W & = & i \bar {Q}_i \tau^a \gamma^{\mu} D_{\nu} Q_j W^{a \mu
\nu}, \nonumber\\ {\cal  O}_B & = & i  \bar {Q}_i \gamma^{\mu} D_{\nu}
Q_j B^{ \mu \nu}, \
\label{op}
\eea where $Q_{i,j}$ are the  left-handed quark doublets, $i,j$ are
the generation indices that refer to the second and third families 
respectively, and
$${\vec     D_{\mu}}     =      {     \vec     \partial}_{\mu}     +ig
A_{\mu}^a\frac{\tau^a}{2} +ig'B_{\mu}\frac{Y}{2}.  $$ Hence we rewrite
Eq.~\ref{eff_lag} as,  \bea {\cal  L}& = &  {\cal L}_{SM}  + \frac{a_W
{\cal O}_W +a_B {\cal O}_B}{\Lambda^2}, \
\label{eff_lag_1}
\eea  As indicated in  the previous  section, the  operators generate
FCNC vertices  with a $q^2$  dependence resulting in new  physics FCNC
effects  in  the  top  sector   that  are  enhanced  by  a  factor  of
$(m_t/m_b)^2$  compared   to  new   physics  effects  in   the  bottom
sector. Such  $q^2$ dependent operators were  previously considered in
the context of single top production\cite{datta-zhang}. One can also write down operators involving the Higgs field which can generate top 
FCNC processes\cite{perez}. Since, the mechanism of electroweak symmetry 
breaking and the Higgs sector of the SM are untested we will not consider 
those operators in our analysis.
Now, before we  go into the details of the  calculations, it is worthwhile
to see how such   interactions might arise. Consider the interaction
involving only the second and third  family quarks of the type: \bea {
\cal  L}_0 &  = &C_3  \bar {Q}_3{  \widetilde {Q}}_3  {\widetilde {X}}
+C_2\bar {Q}_2{ \widetilde {Q}}_2 {\widetilde {X}} + h.c,\
\label{lag_o}
\eea where we have  suppressed any particle indices.  The ${\widetilde
{X}}$ could be a  scalar/pseudoscalar, vector/axial vector etc. and the
${  \widetilde {Q}}_3$  could  be  spin 0  or  spin $  {  1 \over  2}$
objects. Let  us now  suppose there  is mixing such  that in  the mass
basis,  \bea  \wt  {Q}_2  \rightarrow  \wt{Q}_2  \cos{\phi}  -\wt{Q}_3
\sin{\phi},  \nonumber\\  \wt  {Q}_3 \rightarrow  \wt{Q}_2  \sin{\phi}
+\wt{Q}_3 \cos{\phi},\
\label{smix}
\eea  where  $ \phi$  is  the mixing  angle.   One  can then  rewrite,
Eq.~\ref{lag_o} as, \bea  { \cal L}_0 & =  &C_3 \bar {Q}_3{ \widetilde
{Q}}_3 {\widetilde {X}} \cos{\phi}  +C_3 \bar {Q}_3{ \widetilde {Q}}_2
{\widetilde   {X}}  \sin{\phi}   +C_2\bar  {Q}_2{   \widetilde  {Q}}_2
{\widetilde   {X}}\cos{\phi}   -C_2\bar   {Q}_2{   \widetilde   {Q}}_3
{\widetilde {X}}\sin{\phi}\nonumber\\ && + h.c\
\label{lag_final}
\eea  Now we  consider  vertex corrections  involving an  intermediate
$\wt{Q}_3$ or $\wt{Q}_2$  and $\widetilde{X}$.  These corrections will
generate  the   following  vertices:  \bea  \bar{Q}_2   Q_3  V  \equiv
C_2C_3^*\left[ f(\wt  {Q}_2)-f(\wt {Q}_3) \right] \sin{\phi}\cos{\phi}
\nonumber\\   \bar{Q}_3   Q_3    V   \equiv   |C_3|^2   \left[   f(\wt
{Q}_3)\cos^2{\phi}-f(\wt   {Q}_2)  \sin^2{\phi}   \right]  \nonumber\\
\bar{Q}_2 Q_2  V \equiv |C_2|^2  \left[ f(\wt {Q}_2)\cos^2{\phi}-f(\wt
{Q}_3) \sin^2{\phi} \right], \
\label{op_gen} 
\eea 
where $V$ is the $W,Z, \gamma$ and
$f's$ are the loop functions. It is clear that by proper choice of the
parameters one can  make the second operator, $\bar{Q}_3  Q_3 V$,  small
enough  without suppressing  the first  flavor changing  operator. The
second  operator can  contribute to  $Z \to  \bar{b}_L b_L$  where new
physics effects  are strongly constrained \cite{pdg}.  This  is just a
scenario where  the structure  in Eq.~\ref{op} may  be generated.
Since we are adopting a model independent approach, we will not discuss
specific models anymore.

These   operators   in  Eq.~\ref{op}   lead   to  the   following
interactions, \bea {\cal L}_{C}  &= & i \frac{a_W}{\sqrt{2} \Lambda^2}
\left [ \bar {c}  \gamma_\mu(1-\gamma_5)\partial_{\nu} b W^{+ \mu \nu}
+\bar {s} \gamma_\mu(1-\gamma_5) \partial_\nu t W^{- \mu \nu} \right],
\nonumber\\ {\cal L}_{tcZ}  &= & i \frac{a_Wc -  a_B s }{{2\Lambda^2}}
\left  [ \bar  {c} \gamma_\mu(1-\gamma_5)\partial_\nu  t Z^{  \mu \nu}
\right],  \nonumber\\ {\cal  L}_{tc\gamma} &=&  i \frac{a_Ws  +  a_B c
}{{2\Lambda^2}} \left [  \bar {c} \gamma_\mu(1-\gamma_5)\partial_\nu t
A^{ \mu \nu}  \right], \nonumber\\ {\cal L}_{bsZ} &=&  i \frac{-a_Wc -
a_B      s       }{{2\Lambda^2}}      \left      [       \bar      {s}
\gamma_\mu(1-\gamma_5)\partial_\nu b Z^{ \mu \nu} \right], \nonumber\\
{\cal L}_{bs\gamma} &= & i \frac{-a_Ws + a_B c }{{2\Lambda^2}} \left [
\bar {s} \gamma_\mu(1-\gamma_5)\partial_\nu b A^{ \mu \nu} \right],\
\label{lag}
\eea  where  $c  =  \cos{\theta_W}$  and $s  =  \sin{\theta_W}$,  with
$\theta_W$ being the Weinberg angle.

The Lagrangian  above generates  momentum dependent vertices.   We can
combine  the processes  as $t(p)  \to c(k)  V(q)$ and  $b(p)  \to s(k)
V(q)$, where  $V=W,Z, \gamma$.  {}  For $b$ decays the  massive vector
bosons have  to be off-shell.  The vertices for various  processes can
now be written as,
\bea {\cal  L}_{bcW} &= &  - i \frac{a_W}{\sqrt{2} \Lambda^2}  \left [
\bar   {c}  \gamma_\mu(1-\gamma_5)\left(  q^\mu   p^\nu-  q   \cdot  p
\delta^{\mu  \nu}\right)  b   W^{+  \nu}  \right],  \nonumber\\  {\cal
L}_{tsW} &=  & -  i \frac{a_W}{\sqrt{2} \Lambda^2}  \left [  \bar {s}
\gamma_\mu(1-\gamma_5)  \left(  q^\mu  p^\nu-  q \cdot  p  \delta^{\mu
\nu}\right) b  W^{- \nu} \right],  \nonumber\\ {\cal L}_{tcZ} &=  &- i
\frac{a_Wc   -    a_B   s   }{2   \Lambda^2}   \left    [   \bar   {c}
\gamma_\mu(1-\gamma_5)\left(  q^\mu  p^\nu-   q  \cdot  p  \delta^{\mu
\nu}\right)t Z^{\nu} \right], \nonumber\\ {\cal L}_{tc\gamma} & = &- i
\frac{a_Ws   +    a_B   c    }{{2\Lambda^2}}   \left   [    \bar   {c}
\gamma_\mu(1-\gamma_5)\left(  q^\mu  p^\nu-   q  \cdot  p  \delta^{\mu
\nu}\right)t  A^{\nu} \right],  \nonumber\\ {\cal  L}_{bsZ} &=  &  - i
\frac{-a_Wc   -   a_B   s   }{{2   \Lambda^2}}  \left   [   \bar   {s}
\gamma_\mu(1-\gamma_5)\left(  q^\mu  p^\nu-   q  \cdot  p  \delta^{\mu
\nu}\right)b Z^{ \nu} \right], \nonumber\\ {\cal L}_{bs\gamma} &= & -i
\frac{-a_Ws   +   a_B   c   }{{2   \Lambda^2}}  \left   [   \bar   {s}
\gamma_\mu(1-\gamma_5)\left(  q^\mu  p^\nu-   q  \cdot  p  \delta^{\mu
\nu}\right)b A^{\nu} \right].\
\label{ver}
\eea

We now consider the constraints  on the couplings above. We begin with
 $ b \to  s \gamma$.  The SM  amplitude for $b \to s  \gamma$ is given
 by,
\begin{eqnarray}
M_{b\to s \gamma}^{SM}  &= &-V_{tb}V^*_{ts}{G_F\over \sqrt{2}} {e\over
8 \pi^2} C_7(\mu) \bar s \sigma_{\mu\nu}A^{\mu\nu} (m_s L + m_bR)b, \
\label{bsgammaSM}
\eea where $ L(R)  = {( 1 \mp \gamma_5) }$.  Now  we can write the $bs
\gamma$ vertex from Eq.~\ref{lag} as  \bea M_{b\to s \gamma}^{NP} &= &
- \frac{-a_Ws + a_B c }{{4\Lambda^2}} \bar {s} \sigma_{\mu\nu} A^{ \mu
\nu} \left[ m_b R+m_s L\right]b.\
\label{bsgammaNP}
\eea  Comparing  with  the  SM   expression  we  have,  \bea  x&  =  &
 \Big\vert\frac{ { M}_{bs\gamma}^{NP}} {{ M}_{bs\gamma}^{SM}}\Big\vert
 = \left \vert \left[\frac{-a_W  + a_B c/s }{g}\right] \left[ \frac{16
 \pi^2  M_W^2}{\Lambda^2}\right]  \left[  \frac{  1}{g^2V_{tb}V_{ts}^*
 C_7(\mu)}\right] \right \vert . \
\label{x}
\eea   With  $\Lambda   \sim  1   $  TeV,   $C_7(\mu=m_b)   =-0.280  $
\cite{bsgamma}  and $|V_{tb}V_{ts}^*|  =0.04$, we  have $  x  \sim 214
[\frac{-a_W  +  a_B  c/s }{g}]$.   In  other  words  for $  x\sim  1$,
$[\frac{-a_W + a_B c/s}{g}] \sim 0.004$. This difference between $a_W$
and $a_B c/s$ then arises most likely at the loop level. It is interesting to speculate how this scenario might arise in some models of new physics. While we do not present a concrete model, we  refer to Eq.~\ref{lag_o} to Eq.~\ref{op_gen} for an understanding of how the relation between $a_W$ and $a_B$ could arise. If the particles $\widetilde {Q}_{2,3}$ have the same couplings to $W_{\mu}$ and 
$B_{\mu}$ as the SM quarks, resulting from some enhanced symmetry 
, then for the generated operators in Eq.~\ref{op}  we would expect $a_W \propto g$ and $a_B \propto g'$ which could then result in the 
relation between $a_W$ and $a_B$ discussed above. 

 Note that if  $a_W = g$ and $a_B=g'$ then the  NP contribution to $bs
\gamma$ vanishes. Since, due to the weak coupling assumption, we expect $a_W  \sim g$ and $a_B \sim g'$ then
the NP contribution to $b\to s \gamma$ are expected to be small due to
cancellation.   Hence  to avoid  constraints  from  $  b \rightarrow  s
\gamma$ we will choose, \bea \frac{ a_Bc}{a_Ws} & = & 1. \
\label{cond}
\eea  With the  above condition,  we can  now rewrite  the  vertices in
Eq.~\ref{lag}  as,  \bea  {\cal  L}_{C}  &=  &  i  \frac{a_W}{\sqrt{2}
\Lambda^2}  \left [  +\bar {c}  \gamma_\mu(1-\gamma_5)\partial_{\nu} b
W^{+ \mu \nu} +\bar {s} \gamma_\mu(1-\gamma_5) \partial_\nu t W^{- \mu
\nu} \right], \nonumber\\  {\cal L}_{tcZ} &= & i  \frac{a_W(c^2 - s^2)
}{{2c\Lambda^2}} \left [ \bar {c} \gamma_\mu(1-\gamma_5)\partial_\nu t
Z^{ \mu \nu} \right], \nonumber\\ {\cal L}_{tc\gamma} &=& i \frac{a_Ws
}{{\Lambda^2}} \left  [ \bar {c}  \gamma_\mu(1-\gamma_5)\partial_\nu t
A^{   \mu   \nu}   \right],   \nonumber\\   {\cal   L}_{bsZ}   &=&   i
\frac{-a_W}{2c\Lambda^2}        \left         [        \bar        {s}
\gamma_\mu(1-\gamma_5)\partial_\nu b Z^{ \mu \nu} \right]. \
\label{new_lag}
\eea  Hence, all  interactions depend  on the  coupling $a_W$.  We now
estimate  the  effects  of  the  anomalous couplings  on  the  various
vertices. To  be specific we choose $|a_W|$ from $0.5g$ to  $2 g$ and
consider NP effects in the charged  current processes $ t \to s W$ and
$b \to c W$.  We start with the $ t\to s W$ vertex which has the form,
\bea {\cal L}_{tsW} &=& \bar{s} \left[ \gamma_{\mu}(a+b \gamma_5) + ic
\frac{\sigma_{  \mu   \nu}  q^{\nu}}{m_t}  +   id  \frac{\sigma_{  \mu
\nu}\gamma_5q^{\nu}}{m_t}\right]t\epsilon^{*\mu} ,\
\label{tsW}
\eea  with  \bea  a  &  =  &i\frac{a_W}{\sqrt{2}}  \left[\frac{M_W^2}{
2\Lambda^2}\right],\nonumber\\  b   &  =  &-i\frac{a_W}{\sqrt{2}}\left[
\frac{M_W^2}{2    \Lambda^2}\right],    \nonumber\\    c    &    =    &
i\frac{a_W}{\sqrt{2}}           \left[\frac{-(m_t-m_s)m_t}{          2
\Lambda^2}\right],\nonumber\\    d   &   =    &   i\frac{a_W}{\sqrt{2}}
\left[\frac{-(m_t+m_s)m_t}{2 \Lambda^2}\right]. \
\label{abcd_tsW}
\eea The SM piece in this  case is given by, \bea {\cal L}^{SM}_{tsW} &=&
\frac{-ig}{\sqrt{2}}V_{ts}\bar{s}   \left[\gamma_{\mu}(  1-  \gamma_5)
\right]t\epsilon^{*\mu} ,\
\label{tsW_SM}
\eea We can estimate the ratio of  the NP to the SM contribution to $t
\to  sW   $  as,  \bea   r_{tsW}  &  \sim  &   \left\vert  \frac{{\cal
L}^{NP}_{tsW}}        {{\cal        L}^{SM}_{tsW}}\right        \vert=
\left\vert\frac{a_W}{gV_{ts}}  \left[\frac{M_W^2}{  2\Lambda^2}\right]
\right \vert \approx 0.08 \left\vert \frac{a_W}{g}\right\vert,\
\label{ratio}
\eea {}for  $\Lambda=1TeV$ and $|V_{ts}|=0.04$.  In the  above, we have
dropped $c$  and $d$ in  the NP contribution.  We do not  expect their
inclusion to change our estimate by much. Hence allowing $ |a_W|= 0.5g
- 2g$,  $r_{tsW}$ can  be between  4  \% to  16 \%.   Allowing the  NP
contribution  to  add  constructively   to  the  SM  contribution,  the
branching  ratio for $t  \to sW$  is doubled  for $r_{tsW}=\sqrt{2}-1$
which leads to $ a_W \sim 5 g$. The estimate made here is rough and 
a more accurate
calculations of the branching ratio for $ t \to s W$ can be found in the 
next section. The result of the calculation, combined with experimental measurements, validates the use of the assumption $|a_W| \sim g$.

Let us  now turn  to $b\to  cW$: The NP  contribution to  this charged
current is  , \bea {\cal L}_{bcW} &=&  \bar{c} \left[\gamma_{\mu}( a+b
\gamma_5)   +  ic   \frac{\sigma_{   \mu  \nu}   q^{\nu}}{m_b}  +   id
\frac{\sigma_{ \mu \nu}\gamma_5q^{\nu}}{m_b}\right]b\epsilon^{*\mu} ,\
\label{bcW_NP}
\eea  with  \bea  a  & =  &i\frac{a_W}{\sqrt{2}}  \left[\frac{q^2}{  2
\Lambda^2}\right],  \nonumber\\  b  &  =  &-i\frac{a_W}{\sqrt{2}}\left[
\frac{q^2}{    2    \Lambda^2}\right],   \nonumber\\    c    &   =    &
i\frac{a_W}{\sqrt{2}}                      \left[\frac{-(m_b-m_c)m_b}{2
\Lambda^2}\right],\nonumber\\    d   &   =    &   i\frac{a_W}{\sqrt{2}}
\left[\frac{-(m_b+m_c)m_b}{ 2 \Lambda^2}\right]. \
\label{abcd_bcW}
\eea
  The   SM  piece   is  given   by,  \bea   {\cal   L}^{SM}_{bcW}  &=&
\frac{-ig}{\sqrt{2}}V_{cb}\bar{c}   \left[\gamma_{\mu}(  1-  \gamma_5)
\right]b\epsilon^{*\mu} ,\
\label{bcW_SM}
\eea
We can estimate the  ratio of the NP to the SM  contribution to $b \to
cW $ as, \bea r_{bcW}  & \sim & \Big|\frac{{\cal L}^{NP}_{bcW}} {{\cal
L}^{SM}_{bcW}}\Big|=  \left\vert \frac{a_W}{gV_{cb}} \left[\frac{q^2}{
2 \Lambda^2}\right] \right\vert .\
\label{ratiob}
\eea  Using $|a_W|  \sim g$,  $\Lambda=1 $TeV,  $V_{cb}=0.04$,  we find
$r_{bcW}  \stackrel{<}{\sim}  10^{-3}$.   Since  $b \to  cW$ is measured
through  $B$ decays  this NP  correction  will be  masked by  hadronic
uncertainties.

We now turn to FCNC vertices  and start with  the $bsZ$ vertex. This
can  be written  using  Eq.~\ref{new_lag}  as, \bea  { M}_{bsZ}  &=&
\bar{s} \left[ \gamma_{\mu}(a+b \gamma_5) + ic \frac{\sigma_{ \mu \nu}
q^{\nu}}{m_b}         +          id         \frac{\sigma_{         \mu
\nu}q^{\nu}}{m_b}\right]b\epsilon^{*\mu} ,\
\label{bsZ_final}
\eea   with   \bea  a   &   =   &  \frac{-a_W}{2c}   \left[\frac{q^2}{
2\Lambda^2}\right],\nonumber\\    b     &    =    &-    \frac{-a_W}{2c}
\left[\frac{q^2}{   2    \Lambda^2}\right],\nonumber\\   c   &    =   &
\frac{-a_W}{2c}\left[              \frac{-(m_b-m_s)m_b}{             2
\Lambda^2}\right],\nonumber\\     d    &     =     &    \frac{-a_W}{2c}
\left[\frac{-(m_b+m_s)m_b}{2 \Lambda^2}\right].
\label{abcd_bsZ}
\eea We see  that the $bsZ$ couplings are suppressed  by $ \sim {m_b^2
\over \Lambda^2} \sim 2.5 \times 10^{-5}$ which is tiny.  One can look
at this in another way.  As a quick estimate, we can compare the $bsZ$
vertex above with the size of the $bsZ$ in Ref~\cite{ZFCNC}. The $bsZ$
vertex, in the  notation of Ref~\cite{ZFCNC}, is given  by, \bea {\cal
L}_{bsZ}   &=&-\frac{g}{4c}U_{sb}   \bar{s}   \left[   \gamma_{\mu}(1-
\gamma_5) \right]b \epsilon^{*\mu} +h.c. ,\
\label{ZFCNC}
\eea where $|U_{sb}| \sim 0.002$  is obtained using the measured $B_s$
mixing.  Comparing with Eq.~\ref{abcd_bsZ},  we obtain, \bea a_W & \sim
& g U_{sb}\frac{\Lambda^2}{m_b^2}, \
\label{aW}
\eea which leads \bea a_W & \sim & 80 g,\
\label{AWnumber}
\eea for $\Lambda \sim 1 $ TeV  and $m_b \sim 5 $ GeV. We have dropped
$c$ and  $d$ in the  NP contribution which  is reasonable for  a quick
guess estimate  for $a_W$. The  value for $a_W$  in Eq.~\ref{AWnumber}
will  result  in  very  large  effects  in  the  top  sector  that  are
inconsistent  with  experimental  constraints. 
As an example, the branching ratio for $ t \to sW $ will be too large in 
contradiction to experimental results.
   In  our  analysis,  as
indicated earlier,  $a_W \sim  g$ and so  the effect of  the anomalous
couplings  on  the $\bsZ$  are  too  small.   Hence the  operators  in
Eq.~\ref{eff_lag} cannot  generate a $bsZ$ coupling of  the right size
to  explain the  possible  hints of  new  physics in  rare $B$  decays
\cite{ZFCNC}.  Stated  in another way,  any anomalous $bsZ$  vertex of
the correct size must arise from  a mechanism that does not affect the
top sector.   This happens in  models with new  vector-like isosinglet
down-type quarks.

We can now proceed to $\tcV$. We can write the $\tcZ$ vertex as,
\bea  {M}_{tcZ} &=&  \bar{c}  \left[\gamma_{\mu}( a+b  \gamma_5) +  ic
\frac{\sigma_{  \mu   \nu}  q^{\nu}}{m_t}  +   id  \frac{\sigma_{  \mu
\nu}\gamma_5q^{\nu}}{m_t}\right]t\epsilon^{*\mu} ,\
\label{tcZ_final}
\eea  with  \bea a  &  = &\frac{a_W(c^2-s^2)}{2c}  \left[\frac{M_Z^2}{
2\Lambda^2}\right],\nonumber\\   b    &   =   &-\frac{a_W(c^2-s^2)}{2c}
\left[\frac{M_Z^2}{2\Lambda^2}\right],\nonumber\\     c    &     =    &
\frac{a_W(c^2-s^2)}{2c}\left[   \frac{-(m_t-m_c)m_t}{   2   \Lambda^2}
\right],\nonumber\\    d    &    =   &    \frac{a_W(c^2-s^2)}{2c}\left[
\frac{-(m_t+m_c)m_t}{ 2 \Lambda^2}\right].\
\label{abcd_tcZ}
\eea
At  this point,  one  may  worry about  constraints  from meson  mixing
on  $d_id_jZ$ ($d_{i,j}$ are down quarks) vertices generated by the  anomalous couplings above at loop level.  We  show below, that the
size of the $tcZ$ couplings  above is consistent with constraints from
$K$ and $B_{d,s}$ mixings.   {} Following Ref~\cite{han1} we write the
$tcZ$ vertex as,
\begin{eqnarray}
\Delta  {\cal L}^{eff}  = -  \frac{g }{  {2 \cos\theta_W}  }  \, \Big[
\kappa_L^{}  Z^\mu \bar  t \gamma_\mu\Big  (\frac{ 1  -  \gamma_5}{2 }
\Big)  c  +\kappa_R^{}  Z^\mu   \bar  t  \gamma_\mu  \Big(\frac{  1  +
\gamma_5}{2}\Big ) c \Big] + {\it h.c.} \ ,
\label{xin}
\end{eqnarray}
where   $\kappa^{}_{L(R)}$ are  free  parameters
determining  the  strength  of  these anomalous  couplings.   Assuming
$CP$-invariance, $\kappa^{}_{L(R)}$ are real. Comparing the above with
Eq.~\ref{abcd_tcZ}, and  neglecting the terms  $c$ and $d$  we obtain,
\bea     \kappa_L&=&    \frac{a_W}{g}\left[c^2-s^2\right]\frac{M_Z^2}{
\Lambda^2},\nonumber\\ \kappa_R& = & 0.\
\label{kappaL}
\eea Using $\Lambda  = 1$ TeV we find $\kappa_L  \sim 4 \times 10^{-3}
({ a_W \over g})$.

In   Ref~\cite{han1},   the   anomalous   coupling   $\kappa_{L}$   in
Eq.~\ref{xin} was constrained by experimental measurements/bounds on
the   induced   flavor-changing  neutral   couplings   of  the   light
fermions. This was done in the following manner: Integrating the heavy
top quark  out of ${\cal L}^{eff}$ generates  an effective interaction
of the form
\begin{equation}
\tilde{\cal L} = \frac{g}{\cos\theta_W}  a_{i j} \bar {f_i} \gamma^\mu
              \Big( \frac{1 - \gamma_5}{2}\Big ) f_j Z_\mu +{\it h.c.}
              \,
\end{equation}
where $f_i  = b, \ s, \  d$.  Evaluating the one-loop  diagram for the
vertex correction gives
\begin{equation}
a_{ij}  =  \frac{\kappa_L^{}}{  16   \pi^2  }  \frac{m_t^2}{v^2}  \  (
V_{ti}V^*_{cj} + V_{tj}V^*_{ci} ) \ {\rm ln} \frac{\Lambda^2}{m_t^2},
\label{aij}
\end{equation}
where  $V_{ij}$  are  the  elements of  the  Cabbibo-Kobayashi-Maskawa
matrix and $\Lambda$ is a cutoff for the effective Lagrangian.

Now  imposing  constraints on  $a_{ij}$  derived  by studying  several
flavor-changing processes, such as $K_L \rightarrow \bar \mu \mu$, the
$K_L-K_S$ mass  difference, $B_{d,s}^0 - \bar  {B_{d,s}^0}$ mixing,  a
bound on $\kappa_L$ was obtained as \cite{han1},
\begin{equation}
\kappa_L < 5 \times 10^{-2} ,
\label{kllimit}
\end{equation}
%
with  $\Lambda  =$1 TeV  and  $m_t  =$ 171  GeV.   Since  the work  in
Ref~\cite{han1}, $B_s$ mixing  has been measured. One can  estimate $ |
\kappa_L|$,  using   this  new  piece   of  experimental  information.
Comparing  Eq~\ref{ZFCNC} with  Eq~\ref{aij}  we can  write down  \bea
\kappa_L     =    8    \pi^2     \frac{v^2}{m_t^2}\frac{1}{{\rm    ln}
\frac{\Lambda^2}{m_t^2}}|U_{sb}|.\
\label{bs}
\eea This gives  $\kappa_L \sim 9 \times 10^{-2}$  which is consistent
with Eq.~\ref{kllimit}. Using  Eq.~\ref{kappaL} and Eq.~\ref{kllimit} one
obtains $ a_W \sim 10 g$. As  shown in the next section, this will lead
to too large a  branching ratio for $ t \to s W$.  Hence $ a_W \sim g$
is quite  consistent with experimental constraints from  mixing and rare
processes in the down quark sector.

We now move to $ t \rightarrow c \gamma$. The matrix element is
\bea {\cal M}_{tc\gamma} &=& \bar{c} \left[ ic \frac{\sigma_{ \mu \nu}
q^{\nu}}{m_t}         +          id         \frac{\sigma_{         \mu
\nu}q^{\nu}}{m_t}\gamma_5\right]t\epsilon^{*\mu} ,\
\label{tcgamma_final}
\eea   with    \bea   a&=&b=0,   \nonumber\\    c   &   =    &   {a_Ws}
\left[\frac{-(m_t-m_c)m_t}{  2 \Lambda^2}\right],\nonumber\\  d  & =  &
{a_Ws}\left[ \frac{-(m_t+m_c)m_t}{ 2 \Lambda^2}\right].\
\label{abcd_tcgamma}
\eea Again, as before the  above vertex may generate a $bs\gamma$ term
via loop effects. Following Ref~\cite{han2} we write,
\begin{eqnarray}
\Delta {\cal L}^{eff} = \frac{1 }{  \Lambda } \, [ \kappa_{\gamma} e ~
\bar t ~\sigma_{\mu\nu}~ c~ F^{\mu\nu} ] + {\it h.c.} \ ,
\label{Leff}
\end{eqnarray} 
where $F^{\mu\nu}$  is the $U_{em}(1)$  field strength tensor;  $e$ is
the corresponding coupling constant;

Comparing with Eq.~\ref{tcgamma_final}  we obtain, \bea \kappa_{\gamma}
&  \sim  &  \frac{a_W}{g}\frac{m_t}{4   \Lambda}.  \  \eea  This  gives
$|\kappa_\gamma| \sim 4.3 \times 10^{-2}(\frac{a_W}{g})$.

The anomalous  top-quark couplings  $\bar t c  \gamma$ can  modify the
coefficients  of  operators  $O_7$  in the  SM  effective  Hamiltonian for $ b \to s \gamma$
\cite{han2}.  With the value  of $\kappa_\gamma$ above, the corrections
to $b \to s \gamma$  are consistent with the experimental measurements
with $a_W \sim g$.

\section{ Numerical Analysis} In this section we provide the branching ratios for $ t \to c Z,~ t \to c \gamma$ and $t \to s W$. 
The  general form  of  the amplitude  $\CA\(t\rightarrow c+V\)$  where
$V=\gamma$ or  $Z$ is,
 \bea  \CA & =  & \bar u_c\left(  a\,\gamma^\mu +
b\,\gamma^\mu\gamma_5   +  ic\,\sigma^{\mu\nu}{q_\nu   \over   m_t}  +
id\,\sigma^{\mu\nu}\gamma_5 {q_\nu\over m_t}\right)u_t \epsilon^*_\mu,
\
\label{general} 
\eea where $\bar  u_t$, $u_c$ and $\epsilon_\mu$ are  the incoming and
outgoing spinors and the gauge boson polarization vector respectively.
In terms of the coefficient functions the decay widths are,
\bea  \Gamma   (t\rightarrow  c\gamma)   &  =&  \frac{1}{8   \pi}  m_t
\left(|c|^2+|d|^2\right), \nonumber\\  \Gamma (t\rightarrow cZ)  & = &
\frac{1}{16       \pi       m_t}      \left(1-{m_Z^2\over       m_t^2}
\right)\left({m_t^2\over             m_Z^2}-1\right)            \left[
(m_t^2+2m_Z^2)(|a|^2+|b|^2) \right.  \nonumber\\  & & \left.  - 6m_Z^2
Re( a^*c-b^*d)  + m_Z^2({m_Z^2\over m_t^2}+2)(|c|^2+|d|^2)  \right]. \
\eea

The same  formula can  be adapted  to the $  t \to  s W$  process. The
branching  ratios for  $t\rightarrow s  W$, $t\rightarrow  c Z  $, and
$t\rightarrow c \gamma$ processes are defined as, \bea
\label{BR}
 BR_{tsW}  & =  & \frac{\Gamma[t  \to s  W]}{\Gamma[m_t]}, \nonumber\\
BR_{tcZ}  & =  &  \frac{\Gamma[t \to  c Z]}{\Gamma[m_t]},  \nonumber\\
BR_{tc \gamma} & = & \frac{\Gamma[t \to c \gamma]}{\Gamma[m_t]}.  \eea
{} For  the top  width we   use $
\Gamma(m_t) \approx \Gamma (t\rightarrow bW) $ which is given by, \bea
\Gamma  (t\rightarrow bW)  &  = &  {G_F\over 8\pi\sqrt{2}}  |V_{tb}|^2
m_t^3     \left(1-{m_W^2\over     m_t^2}\right)    \left(1+{m_W^2\over
m_t^2}-2{m_W^4\over m_t^4}\right) .\
\label{top_width}
\eea For  the   charged current  pieces we have  to include  the SM
contributions also.   For the  $b \to cW$  transition we  have already
shown the NP  contribution to be small and so we  will not consider it
any further. For the  rare decays $\tcV$, since the SM contributions
are tiny we can ignore the  SM terms.  \par In the numerical analysis,
we used the quark masses $m_t = 171.3$ GeV, $m_b = 4.2$ GeV \cite{pdg},
 and  CKM  matrix elements  $|V_{ts}|=0.04042$,  $|V_{tb}|=0.999146$
\cite{ckmfit}.  We  plotted the  branching ratios for  $t\rightarrow c
\gamma$, and $t\rightarrow c Z $  as a function of $ |a_W|$ for $\Lambda
=1$ TeV in Figs.~\ref{Brspl1}(a) and  1(b) respectively.  Here $|a_W|$ is
varied between 0.5 g and 2 g. Also,  the branching ratio for
$t\rightarrow s  W$ is plotted as a  function of $ |a_W|$ for $\Lambda =1$  TeV in
Fig.~\ref{Brspl2}.   The  NP  contributions  added  constructively  and
destructively    to    the    SM    contributions   are    shown    in
Figs.~\ref{Brspl2}(a) and  \ref{Brspl2}(b) respectively. We calculated
the   branching   ratios  $BR_{tsW}   \approx   10.3  \times   10^{-3}
~\mathrm{(constructive)}, ~\approx   3.8  \times  10^{-3}
~\mathrm{(destructive)}$,  $BR_{tcZ}\approx 0.93 \times  10^{-4}$, and
$BR_{tc \gamma} \approx 2.0 \times 10^{-4}$ at $ |a_W|=g$. The branching
ratios for $\tcZ$ and $\tcgam$ are  within the reach of  LHC. Using the
maximum $BR_{tsW}$ above, we can compute, \bea r_t & = & \frac{\Gamma[t
\to bW]}{\sum_{q=d,s,b} \Gamma[t \to Wq]} \sim 0.99.\
\label{rt}
\eea
The experimental measurements give $r_t =0.99^{+0.09}_{-0.08}$ \cite{pdg}, 
which compared to $r_t$ in Eq.~\ref{rt} validates the weak 
coupling assumption $ |a_W| \sim g$.
 
\begin{figure}[htb!]
  \includegraphics[width=7.5cm]{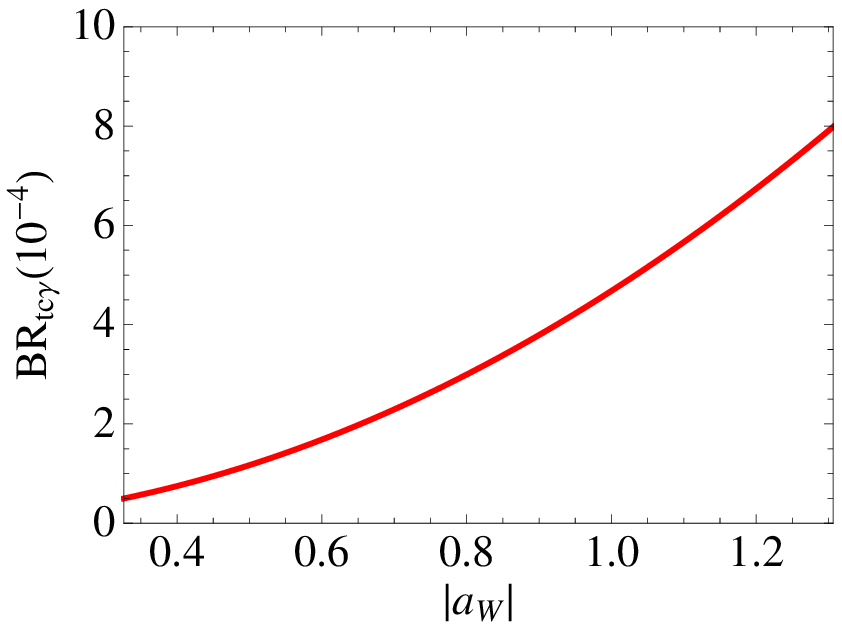}
  \includegraphics[width=7.5cm]{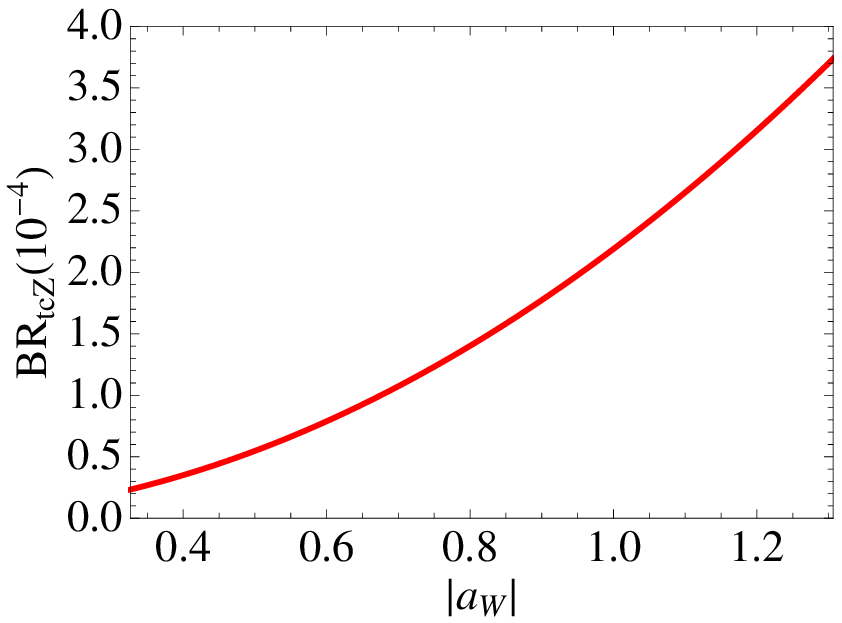}
   \caption{ The branching fractions  (a) $BR_{tc\gamma} (10^{-4})$, and (b) $BR_{tcZ}(10^{-4})$  plotted as a function of $|a_W|$ for $m_t =171.3$ GeV, and $\Lambda = 1$ TeV.  Here  $|a_W|$ is varied between 0.5 g and 2 g.}\label{Brspl1}
\end{figure}
\begin{figure}[htb!]
\centering
  \includegraphics[width=7.0cm]{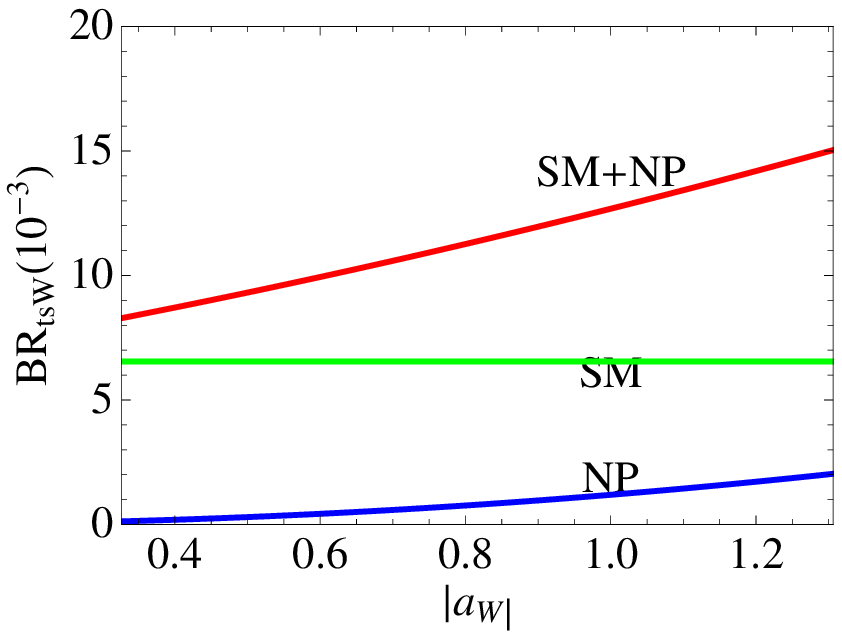}
  \includegraphics[width=7.0cm]{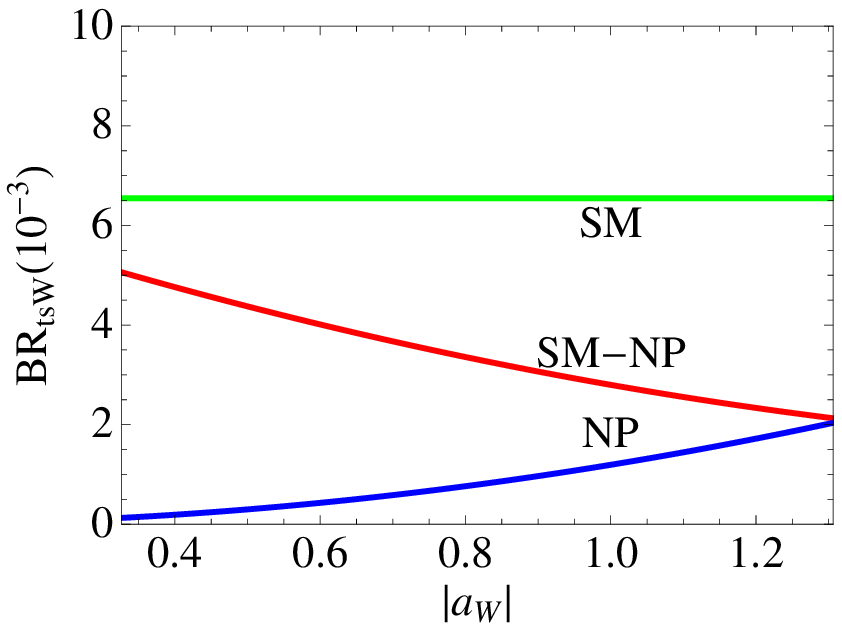}
    \caption{The branching fraction  $BR_{tsW}$ plotted as a function of $|a_W|$ for $m_t =171.3$ GeV, and $\Lambda = 1$ TeV.   Here $|a_W|$ is varied between 0.5 g and 2 g. (a) We assume constructive interference between the SM and NP
   contributions. (b) We assume destructive interference between the SM and NP contributions.}\label{Brspl2}
\end{figure}


\section{Conclusion} In  this paper, we considered rare $\bsV$ and $\tcV$ 
decays that
 arise from the  same non-SM physics,  or in  other words,  the same
higher dimensional operator corrections to the standard model. The existing
constraints from $B$ physics  strongly constrain the NP contributions to
$ t \to c Z(\gamma)$.   In certain situation, the constraints from $B$
decays as well as top branching fraction measurements  still allow   
 branching ratios  
 for $ t \to c Z(\gamma)$ that may be accessible at the LHC.

\section{ Acknowledgments}
The work of A.D was supported by a research grant from the College of Liberal Arts, 
University of Mississippi.

\end{document}